\documentstyle[12pt]{article} 
\normalbaselines
\begin{document}
\bibliographystyle{unsrt} 

\vbox {\vspace{6mm}} 

\begin{center}
{\large \bf CORRELATED LIGHT AND SCHR\"ODINGER CATS}\\[7mm]
V. I. Man'ko\\ 
{\it Lebedev Physical Institute\\ 
53 Leninsky Prospekt, Moscow 117333, Russia}\\[5mm]

\end{center}

\vspace{2mm}

\begin{abstract}

The  Schr\"odinger cat male and female states are discussed. 
The Wigner and Q--functions of 
generalized correlated light are given. 
Linear transformator of photon statistics is reviewed.

\end{abstract}

\section{Introduction}
 
The integral of motion which is quadratic in position and momentum was 
found for classical oscillator 
with time-dependent frequency 
by Ermakov \cite{Er1880}. Two time-dependent 
integrals of motion which are 
linear forms in position and momentum for the classical and quantum 
oscillator with time-dependent frequency were found in \cite{mal70}; 
for a charge moving in varying in time uniform magnetic field, this was 
done in \cite{MalTri69}. For the multimode 
nonstationary oscillatory systems, 
such new integrals of motion, both of Ermakov's 
type (quadratic in positions and momenta) and linear in position and 
momenta, generalizing the results of \cite{mal70} were constructed 
in \cite{MalManTri73}. 
We will consider below the parametric 
oscillator using the integrals of motion.
The Wigner function 
of multimode squeezed 
light is studied 
using such special functions as multivariable 
Hermite polynomials. 

The theory of parametric oscillator is appropriate to 
consider the problem of creation of photons from vacuum in a 
resonator with moving walls (with moving mirrors) which is the 
phenomenon based on the existence of Casimir forces (so-called 
nonstationary Casimir effect). The resonator with moving boundaries 
(moving mirrors, media with time-dependent refractive index) 
produces also effect of squeezing in the light quadratures. In the high 
energy physics very fast particle collisions may produce new types of 
states of boson fields (pions, for example) which are squeezed and 
correlated states studied in quantum optics but almost unknown in particle 
physics, both theoretically and experimentally.

\section{Multimode Quadratic Systems}

The generic nonstationary linear system has the Hamiltonian
\begin{equation}\label{gq1}
H=\frac {1}{2}{\bf Q}B(t){\bf Q}+{\bf C}(t){\bf Q},
\end{equation}
where we use 2N--vectors $~{\bf Q}=~(p_{1},~p_{2},\ldots ,
p_{N},~q_{1},~q_{2},\ldots ,q_{N})$ and $~{\bf C}(t),$ as
well as 2N$\times $2N--matrix $~B(t)$, the Planck constant
$~\hbar =1.$ This system has 2N linear integrals of motion
\cite{dod89}, \cite{mal79} which may be written in vector form
\begin{equation}\label{gq2}
{\bf Q}_{0}(t)=\Lambda (t){\bf Q}+{\bf \Delta }
(t).
\end{equation}
The real symplectic matrix $~\Lambda (t)~$ is the solution to 
the system of equations
\begin{eqnarray}\label{gq3}
\dot \Lambda (t)&=&\Lambda (t)\Sigma B(t),\nonumber\\
\Lambda (0)&=&1,
\end{eqnarray}
where the real antisymmetrical matrix  $~\Sigma ~$ is 2N--dimensional
analog of the Pauli matrix $~i\sigma _{y},$ and the vector 
$~{\bf \Delta }(t)~$ is the solution to the system of equations
\begin{eqnarray}\label{gq4}
\dot {\bf \Delta }(t)&=&\Lambda (t)\Sigma {\bf C}(t),\nonumber\\
{\bf \Delta }(0)&=&0.
\end{eqnarray}
If for time $~t=~0,$ one has the initial Wigner function of the system
in the form
\begin{equation}\label{gq5}
W({\bf p},{\bf q},t=0)=W_{0}({\bf Q}),
\end{equation}
the Wigner function of the 
system at time $~t~$ 
is (due to the density operator  is the integral of motion)
\begin{equation}\label{gq6}
W({\bf p},{\bf q},t)=W_{0}[\Lambda (t){\bf Q}+{\bf \Delta }(t)].
\end{equation}
This formula may be 
interpreted as transformation of 
input  Wigner function 
into output Wigner 
function due to 
symplectic quadrature 
transform (\ref{gq2}). 
An optical 
linear transformator of photon distribution function using this 
output Wigner function is suggested in \cite{polyn}. 

The Hamiltonian (\ref{gq1}) may be rewritten in terms of creation 
and annihilation operators
\begin{equation}\label{gq7}
H=\frac {1}{2}{\bf A}D(t){\bf A}
+{\bf E}(t){\bf A},
\end{equation}
where we use 2N--vectors $~{\bf A}=~(~a_{1},~a_{2},\ldots ,
~a_{N},~a_{1}\dag ,~a_{2}\dag ,\ldots ,~a_{N}\dag )$ and 
$~{\bf E}(t),$ as well as 2N$\times $2N--matrix $~D(t).$
This system has 2N linear integrals of motion \cite{dod89},
\cite{mal79} which are written in vector form
\begin{equation}\label{gq8}
{\bf A}_{0}(t)=M(t){\bf A}+{\bf N}(t).
\end{equation}
The complex matrix $~M(t)~$ is the solution to the system of
equations
\begin{eqnarray}\label{gq9}
\dot M(t)&=&M(t)\sigma D(t),\nonumber\\
M(0)&=&1,
\end{eqnarray}
where the imaginary antisymmetric matrix $~\sigma ~$ is 
2N$\times $2N--analog of the Pauli matrix $~-\sigma _{y},$ and
the vector $~{\bf N}(t)~$ is the solution to the system of equations
\begin{eqnarray}\label{gq10}
\dot {\bf N}(t)&=&M(t)\sigma {\bf E}(t),\nonumber\\
{\bf N}(0)&=&0.
\end{eqnarray}
Analogously to the Wigner function evolution, if for time $~t=~0,$ 
one has the initial Q--function of the system in the form
\begin{equation}\label{gq11}
Q(\alpha ,  \alpha ^{*},t=0)
=Q_{0}({\cal A}),~~~~~{\cal A}=(\alpha, \alpha ^{*}),
\end{equation}
the Q--function of the system at time $~t~$ is
\begin{equation}\label{gq12}
Q(\alpha , \alpha ^{*},t)=
Q_{0}[M(t){\cal A}+{\bf N}(t)].
\end{equation}
Here $~ \alpha =~(~{\bf q}~+~i{\bf p})/\sqrt 2.$

For time-independent Hamiltonian (\ref{gq1}), the matrix 
$~\Lambda (t)~$ is 
\begin{equation}\label{gq13}
\Lambda (t)=\exp (\Sigma Bt),
\end{equation}
and the vector $~{\bf \Delta }(t)~$ is
\begin{equation}\label{gq14}
{\bf \Delta }(t)
=\int _{0}^{t}\exp (\Sigma B\tau )~\Sigma ~{\bf C}(\tau )~d\tau .
\end{equation}
For time-independent Hamiltonian (\ref{gq7}), the matrix $~M(t)~$ is 
\begin{equation}\label{gq15}
M(t)=\exp (\sigma Dt),
\end{equation}
and the vector $~{\bf N}(t)~$ is
\begin{equation}\label{gq16}
{\bf N}(t)
=\int _{0}^{t}\exp (\sigma D\tau )~\sigma ~{\bf E}(\tau )~d\tau .
\end{equation}
For time-dependent linear systems, the Wigner function of generic squeezed 
and correlated state (generalized correlated state \cite{sudar} ) has 
Gaussian form and it was calculated in \cite{dod89}.

Thus the evolution of the Wigner function and Q--function for 
systems  with quadratic Hamiltonians for any state is given by 
the following prescription. Given the Wigner function $~W({\bf p,q},t=0)$ 
for the initial moment of time $~t=0.$ Then the Wigner function for time  
$~t$ is obtained by the replacement
$$W({\bf p,q},t)=~W({\bf p}(t),~{\bf q}(t),~t=0),$$
where time-dependent arguments are the linear integrals of motion
of the quadratic system found in \cite{dod89}, \cite{MalManTri73}, 
and \cite{tri73}. This formula was given as integral with 
$~\delta $--function kernel in \cite{vol191}. The linear integrals of 
motion describe initial values of classical trajectories in 
the phase space of the system. The same ansatz is used for the Q--function. 
Namely, given the Q--function of the quadratic system $~Q({\bf B},~t=0)$  
for the initial moment of time $~t=0.$ Then the Q--function for time $~t$ 
is given by the replacement
$$Q({\bf B},~t)=~Q({\bf B}(t),~t=0),$$
where the 2N--vector $~{\bf B}(t)$ is the integral of motion linear in the 
annihilation and creation operators.
This ansatz follows from the statement that the density operator of the
Hamiltonian system is the integral of motion, and its matrix elements in any 
basis must depend on appropriate integrals of motion.

\section{Multimode Mixed Correlated Light}
 
The most general mixed squeezed state of the N--mode light with a
Gaussian density operator $~\hat{\rho }$ is described by 
the Wigner function $~W({\bf p},{\bf q})$ of the generic Gaussian form, 
\begin{equation}
W({\bf p},{\bf q})=\det {\bf M}^{-\frac{1}{2}}\exp\left
[-\frac 12({\bf Q}-<{\bf Q}>){\bf M}^{-1}({\bf Q}-<{\bf Q}>)\right],
\label{m1}
\end{equation}
where 2N parameters $~<p_i>$ and $~<q_i>$, $~i=1,2,\ldots ,N$, combined into 
vector $~<{\bf Q}{\bf >}$, are average values of 
quadratures,
\begin{eqnarray}
<{\bf p}>&=&\mbox{Tr}~\hat{\rho}\hat {{\bf p}},\nonumber\\
<{\bf q}>&=&\mbox{Tr}~\hat{\rho}\hat {{\bf q}}.
\label{m3}
\end{eqnarray}
A real symmetric dispersion matrix $~{\bf M}$ consists of 2N$^2$+N
variances
\begin{equation}
{\cal M}_{\alpha\beta}=\frac {1}{2}<\hat Q_{\alpha}\hat
Q_{\beta}+\hat Q_{\beta}\hat Q_{\alpha}> -<
\hat Q_{\alpha}><\hat Q_{\beta}>
,~~~~~~~~~~~\alpha ,\beta =1,2,\ldots ,2N.
\label{m4}
\end{equation}
They obey uncertainty relations constraints \cite{dod89}. According to 
previous section the Wigner function of parametric linear system with 
initial value (\ref{m1}) is 
\begin{equation}
W({\bf p},{\bf q},t)=\det {\bf M}^{-\frac{1}{2}}\exp\left
[-\frac 12(\Lambda (t){\bf Q}+
{\bf \Delta }(t)-<{\bf Q}>){\bf M}^{-1}(\Lambda (t){\bf Q}+{\bf \Delta } (t)-
<{\bf Q}>)\right],
\label{wt}
\end{equation}
The photon distribution function of the state  (\ref{m1})
\begin{equation}
{\cal P}_{{\bf n}}=\mbox{Tr}~\hat{\rho }|{\bf n}><{\bf n}|,
~~~~~~~{\bf n}=(n_1,n_2,\ldots ,n_N),
\label{m5}
\end{equation}
where the state $~|{\bf n}>$ is photon number state, which was 
calculated in \cite{olga}, \cite{semen} and it is
\begin{equation}
{\cal P}_{{\bf n}}={\cal P}_0\frac {H_{{\bf n}{\bf n}}^{
\{{\bf R}\}}({\bf y})}{{\bf n}!}.
\label{m6}
\end{equation}
The trace (\ref{m5}) may be calculated using the explicit form of the
Wigner function of the operator $~|{\bf m}><{\bf n}|$ 
(see, \cite{dod89}) which is the product of Wigner functions of 
one-dimensional oscillator expressed in terms of Laguerre polynomials
of the form
\begin{equation}\label{wigfunodo}
W_{mn}(p,q)=2^{m-n+1}(-1)^{n}\sqrt {\frac {n!}{m!}}
\left (\frac {q-ip}{\sqrt 2}\right )^{m-n}e^{-(p^{2}+q^{2})}
L_{n}^{m-n}\left (2(q^{2}+p^{2})\right ).
\end{equation}
The function $~H_{{\bf n}{\bf n}}^{\{{\bf R}\}}({\bf y})$ is 
multidimensional Hermite polynomial. The probability to have no 
photons is
\begin{equation}
{\cal P}_0=\left[\det\left({\bf M}+\frac 12{\bf I}_{2N}\right)\right
]^{-1/2}\exp\left[-<{\bf Q}>\left(2{\bf M}+{\bf I}_{2N}\right
)^{-1}<{\bf Q}>\right],
\label{m7}
\end{equation}
where we introduced the matrix
\begin{equation}
{\bf R}=2{\bf U}^{\dag }(1+2{\bf M})^{-1}{\bf U}^{*}-\sigma _{Nx},
\label{m8}
\end{equation}
and the matrix
\begin{equation}
\sigma _{Nx}=\left(\begin{array}{cc}
0&{\bf I}_N\\
{\bf I}_N&0\end{array}\right).
\label{m9}
\end{equation}
The argument of Hermite polynomial is
\begin{equation}
{\bf y}=2{\bf U}^t({\bf I}_{2N}-2{\bf M})^{-1}<{\bf Q}>,
\label{m10}
\end{equation}
and the 2N--dimensional unitary matrix
\begin{equation}
{\bf U}=\frac 1{\sqrt {2}}\left(\begin{array}{cc}
-i{\bf I}_N&i{\bf I}_N\\
{\bf I}_N&{\bf I}_N\end{array}\right)
\label{m11}
\end{equation}
is introduced, in which $~{\bf I}_N$ is the N$\times $N--identity 
matrix. Also, we use the notation
$${\bf n}!=n_1!n_2!\cdots n_N!.$$

The mean photon number for j--th mode is expressed in terms of
photon quadrature means and dispersions
\begin{eqnarray}
<n_j>=\frac 12(\sigma_{p_jp_j}+\sigma_{q_jq_j}-1)
+\frac 12(<p_j>^2+<q_j>^2).
\label{m12}
\end{eqnarray}
The photon distribution function for transformed state (\ref{wt}) is given 
by the same formulae (\ref{m6}), (\ref{m7})--(\ref{m11}) but with 
changed dispersion matrix
\begin{equation}
\widetilde {\bf M}=\Lambda ^{-1}{\bf M}\Lambda ^{-1t},
\label{dis}
\end{equation} 
and quadrature means 
\begin{equation}
<\widetilde {\bf Q}>=\Lambda ^{-1}({\bf \Delta }-<{\bf Q}>).
\label{means}
\end{equation} 
Thus we have a linear transformator of photon statistics suggested 
in \cite{polyn}.

Let us now introduce a complex 2N--vector $~{\bf B}=(\beta _{1},~\beta _{2},
~\ldots ,~\beta _{N},~\beta _{1}^{*},~\beta _{2}^{*},~\ldots ,
~\beta _{N}^{*})$. Then the Q--function  is
the diagonal matrix element of the density operator in coherent
state basis $~|~\beta _{1},~\beta _{2},~\ldots ,~\beta _{N}>.$
This function is the generating function for matrix elements of the
density operator in the Fock basis $~|${\bf n}$>$ which has been calculated 
in \cite{semen}. In notations corresponding to the Wigner function (\ref{m1}) 
the Q--function is
\begin{equation}
Q({\bf B})={\cal P}_{0}\exp \left [-\frac {1}{2}{\bf B}(R+\sigma _{Nx}){\bf B}+
{\bf B}R{\bf y}\right ].
\label{m13}
\end{equation}
Thus, if the Wigner function (\ref{m1}) is given one has the Q--function.
Also, if one has the Q--function (\ref{m13}), i.e., the matrix $~R$ and 
vector {\bf y}, the Wigner function may be obtained due to 
relations
\begin{eqnarray}
{\bf M}&=&{\bf U}^{*}(R+\sigma _{Nx})^{-1}{\bf U}^{\dag }
-1/2,\nonumber\\
<{\bf Q}>&=&{\bf U}^{*}[1-(R+\sigma _{Nx})^{-1}\sigma _{Nx}]{\bf y}.
\label{m14}
\end{eqnarray}
Multivariable Hermite polynomials describe the photon distribution 
function for the multimode mixed and pure correlated light \cite{olga},
\cite{md94}, \cite{dodon94}. The nonclassical state of light may be 
created due to nonstationary Casimir effect \cite{jslr}, \cite{dkm} and the 
multimode oscillator is the model to describe the behaviour of 
squeezed and correlated photons.

\section{Parametric Oscillator}

For the parametric oscillator with the Hamiltonian 
\begin{equation}
H=-\frac {\partial ^{2}}{2\partial x^{2}}
+\frac {\omega ^{2}(t)x^{2}}{2},
\label{s60}
\end{equation}
where we take $~\hbar=m=\omega (0)=1$, there exists the 
time-dependent integral of motion found in \cite {mal70}
\begin{equation}
A=\frac {i}{\sqrt 2}[\varepsilon (t)p-\dot \varepsilon (t)x],
\label{s61}
\end{equation}
where
\begin{equation}
\ddot \varepsilon (t)+\omega ^{2}(t)\varepsilon (t)=0,~~~~~~
\varepsilon (0)=1,~~~~~~\dot \varepsilon (0)=i,
\label{s62}
\end{equation}
satisfying the commutation relation
\begin{equation}
[A,~A\dag ]=1.
\label{s63}
\end{equation}
It is easy to show that packet solutions of the Schr\"odinger equation
may be introduced and interpreted as coherent states \cite{mal70}, since  
they are eigenstates of the operator $~A$ (\ref{s61}), of the form
\begin{equation}
\Psi _{\alpha }(x,t)=\Psi _{0}(x,t)\exp \left (-\frac {|\alpha |^{2}}{2}-
\frac {\alpha ^{2}\varepsilon ^{*}(t)}{2\varepsilon (t)}
+\frac {{\sqrt 2}\alpha x}{\varepsilon}\right ),
\label{s64}
\end{equation}
where
\begin{equation}
\Psi _{0}(x,t)=\pi ^{-1/4}\varepsilon (t)^{-1/2}
\exp \frac {i\dot \varepsilon (t)x^{2}}{2\varepsilon (t)}
\label{s65}
\end{equation}
is analog of the ground state of the oscillator and $~\alpha $ is a
complex number. 

Variances of the position and momentum of the
parametric oscillator in the state (\ref{s64}), (\ref{s65}) are
\begin{equation}
\sigma _{x}=\frac {|\varepsilon (t)|^{2}}{2},~~~~~~\sigma _{p}
=\frac {|\dot \varepsilon (t)|^{2}}{2},
\label{s66}
\end{equation}
and the correlation coefficient $~r$ of the position and momentum has
the value corresponding to minimization of the Schr\"odinger uncertainty
relation \cite{schrod}
\begin{equation}
\sigma _{x}\sigma_{p}=\frac {1}{4}\frac {1}{1-r^{2}}.
\label{s67}
\end{equation}
If $~\sigma _{x}<1/2~~(\sigma _{p}<1/2)$ we have squeezing in photon 
quadrature components.

The analogs of orthogonal and complete system of states which are excited 
states of stationary oscillator are obtained by expansion of (\ref{s64})
into power series  in $~\alpha .$ We have
\begin{equation}\label{insert1}
\Psi _{m}(x,t)=\left (\frac {\varepsilon ^{*}(t)}{2\varepsilon (t)}
\right )^{m/2}\frac {1}{\sqrt {m!}}\Psi _{0}(x,t)H_{m}\left (\frac {x}
{|\varepsilon (t)|}\right ),
\end{equation}
and these squeezed and correlated number states are eigenstates of 
invariant $~A^{\dag }A.$ In case of periodical dependence of 
frequency on time the classical solution in stable regime may be taken in 
Floquet form
\begin{equation}
\varepsilon (t)=e^{i\kappa t}u(t),
\label{Fl}
\end{equation}
where $~u(t)$ is a periodical function of time. Then the states 
(\ref{insert1}) are quasienergy states realizing the unitary irreducible
representation of time translation symmetry group of the Hamiltonian
and the parameter $~\kappa $ determines the quasienergy spectrum.
Unstable classical solutions give continuous spectrum of quasienergy states.

The partial cases of parametric oscillator are free motion 
$~(~\omega (t)=0~),$ stationary harmonic oscillator 
$~(~\omega ^{2}(t)=1~),$ and repulsive oscillator
$~(~\omega ^{2}(t)=-1~).$ The solutions obtained above are described by 
the function $~\varepsilon (t)~$ which is equal to 
$~\varepsilon (t)=1+it,$ for free particle, $~\varepsilon (t)=e^{it},$ 
for usual oscillator, and $~\varepsilon (t)=\cosh t+i\sinh t,$ for
repulsive oscillator.

Another normalized solution to the Schr\"odinger equation 
\begin{equation}
\Psi _{\alpha m}(x,t)=2N_{m}\Psi _{0}(x,t)\exp \left (-\frac {|\alpha |^{2}}
{2}-\frac {\varepsilon ^{*}(t)\alpha ^{2}}{2\varepsilon (t)}\right )\cosh
\frac {{\sqrt 2}\alpha x}{\varepsilon (t)},
\label{s68}
\end{equation}
where
\begin{equation}
N_{m}=\frac {\exp (|\alpha |^{2}/2)}{2\sqrt {\cosh |\alpha |^{2}}},
\label{s69}
\end{equation}
is the even coherent state \cite{dod74} (the Schr\"odinger cat male 
state). The odd coherent state of the parametric oscillator 
(the Schr\"odinger cat female state)
\begin{equation}
\Psi _{\alpha f}(x,t)=2N_{f}\Psi _{0}(x,t)\exp \left (-\frac {|\alpha |^{2}}
{2}-\frac {\varepsilon ^{*}(t)\alpha ^{2}}{2\varepsilon (t)}\right )\sinh
\frac {\sqrt {2}\alpha x}{\varepsilon (t)},
\label{s70}
\end{equation}
where
\begin{equation}
N_{f}=\frac {\exp (|\alpha |^{2}/2)}{2\sqrt {\sinh |\alpha |^{2}}},
\label{s71}
\end{equation}
satisfies the Schr\"odinger equation and is the eigenstate of the 
integral of motion $~A^{2}$ (as well as the even coherent state) 
with the eigenvalue $~\alpha ^{2}$. These states are one-mode examples of 
squeezed and correlated Schr\"odinger cat states constructed in \cite{nikon}. 
The experimental creation of the Schr\"odinger cat states is discussed in 
\cite{Har}. These states belong to family of nonclassical 
superposition states studied in  \cite{nieto}, \cite{janszky}.

\end{document}